\documentclass[11pt,a4paper]{article}
\usepackage[utf8]{inputenc}
\usepackage{amsmath}
\usepackage{amsfonts}
\usepackage{amssymb}
\usepackage{graphicx}
\usepackage{bm}
\usepackage{booktabs}       
\usepackage{multirow}
\usepackage{epstopdf}
\usepackage{hyperref}       
\usepackage{url}            
\usepackage[top=2.52cm, bottom=2.52cm, left=2.52cm, right=2.52cm]{geometry}
\usepackage{xcolor} 
\usepackage{setspace}
\begin{document}
\title{\textbf{Experimental observation of Berreman modes in uniaxial anisotropic nanoporous alumina film on aluminium substrate}}



\author{Dheeraj Pratap\footnote{Corresponding author's email: dheeraj.pratap@csio.res.in} $^{, a, b}$, Jitendra Kumar Pradhan\footnote{Corresponding author's email: jkpradhan99@gmail.com} $^{, a, c}$, \\ and Subramanium Anantha Ramakrishna$^{a, d}$ \\ \\
$^{a}$Department of Physics, Indian Institute of Technology Kanpur,  Kanpur-208016, India\\
	$^{b}$Biomedical Applications, CSIR-Central Scientific Instruments Organization, \\ Sector-30C, Chandigarh-160030\\
	$^{c}$Department of Physics, Rajendra University, Balangir-767002, India\\
	$^{d}$CSIR-Central Scientific Instruments Organization, Sector-30C, Chandigarh-160030\\
}

\date{}

\maketitle

\begin{abstract}
In this article, we demonstrate experimentally and verified numerically the excitation of Berreman modes that propagate in a dielectric film of uniaxial anisotropic nanoporous alumina grown on an aluminium substrate. It is an air-dielectric-metal asymmetric polaritonic system with a real part of the effective permittivity having a value near zero. The modes are excited at a wavelength lower than the epsilon near zero wavelength region. Minimum reflection is observed for the mid-infrared p-polarized light, while maximum reflection is observed for the s-polarized light. The experimental results are numerically reproduced for both p- and s-polarized light and confirm the Berreman modes excitation in the system. At the exciting wavelength, the field is confined in the dielectric region near the air-dielectric interface. The reported system is straightforward and could be easily fabricated over a large scale and is helpful in a variety of mid-infrared applications such as thermal management systems, sensors, passive radiative cooling devices, non-linear applications, and terahertz frequency generation.

\vspace{3mm}
\end{abstract}


\section{Introduction}
In the field of nanophotonics, the infrared (IR) characteristics of polar dielectric materials have attracted considerable interest in recent years~\cite{dai2014tunable,caldwell2015low,low2017polaritons}. Reststrahlen bands exist between the spectral regions of such polar dielectrics' transverse optical and longitudinal optical phonons.  The real part of the dielectric function of such phononic materials is negative in the Reststrahlen band rendering a high reflectivity of light within this band~\cite{bohren2008absorption,adachi2012optical}. The polar materials can sustain phonon polariton modes within the Reststrahlen band, which are analogous to plasmon polaritons~\cite{maksimov1984modern,shen2009surface,caldwell2014sub,feng2015localized}. Since the phonon polaritons rely on phononic rather than electronic dynamics, their scattering lifetimes are orders of magnitude longer than plasmonic modes, resulting in phonon polariton resonances with substantially smaller line widths and reduced losses~\cite{chen2014spectral,xu2014one}. These low-loss phonon polariton modes have been discovered as interesting candidates for a variety of applications, including customised chemical sensing~\cite{berte2018sub}, thermal emission~\cite{wang2017phonon,sakotic2020berreman}, and chip-level optical components~\cite{dai2015subdiffractional}, among others. Near the longitudinal optical phonon frequency in the spectral range where the real part of the permittivity, epsilon, is near zero, (ENZ) a particularly intriguing class of phonon polariton modes occur in dielectric thin films~\cite{berreman1963infrared,vassant2012berreman,vassant2012epsilon}. Guided modes, where light is reflected at the top and bottom interfaces of a dielectric sub-wavelength thin film, can be supported in the ENZ spectral region for a dielectric sub-wavelength thin film. The polariton mode is a leaky guided mode that is referred to as a Berreman mode on the low-momentum side of the light line, and the mode is evanescently restricted and is known as an ENZ polariton mode on the high-momentum side of the light line~\cite{vassant2012berreman,remesh2021exploring}. Berreman and ENZ polaritons, like localized surface phonon polariton (SPhP) resonances, create strongly restricted electric fields, enhancing light-matter interaction~\cite{passler2019second}. Such polariton modes can have an impact on a wide range of essential applications, including absorption and thermal emission management~\cite{vassant2013electrical,sakotic2020berreman}, IR directional absorption~\cite{luk2014directional}, and better nonlinear processes~\cite{passler2019second}. ENZ polaritons have recently been demonstrated to strongly couple to other SPhPs in multilayer structures, opening up the possibility of highly controllable SPhP modes in complex hybrid devices~\cite{passler2018strong}. Mid-IR nano spectroscopy of Berreman mode and ENZ local field confinement in a thin film of SiO$_2$ on Si was reported by Shaykhutdinov et al.~\cite{shaykhutdinov2017mid}. The second harmonic generation from the Berreman mode was observed in AlN and SiC system~\cite{passler2019second}. In the ITO/HfO$_2$/Au system, field-Effect tunable and broadband ENZ perfect absorbers were reported by Anopchenko et al.~\cite{anopchenko2018field}. Recently, the broadband terahertz generation using the ITO film on glass was reported by Jia et al.~\cite{jia2021broadband}. Nanoporous alumina of various thicknesses has been used to make the passive radiative cooling surfaces~\cite{fu2019daytime,pradhan2021thin}. However, there is no report on the excitation of Berreman modes in the nanoporous alumina on the aluminium system. The sub-wavelength sized diameter of the unidirectional nanopores in the nanoporous alumina create uniaxial anisotropy. This anisotropy is geometrically induced in the nanoporous alumina. 

In this article, we experimentally demonstrate the excitation of polaritonic modes in air-filled uniaxial anisotropic nanoporous alumina grown on an aluminium substrate at the mid-IR wavelength regime. These modes are characterized as Berreman modes that are excited at the air-nanoporous alumina interface. We have also verified the experimental results with computational modelling. The organization of the paper is as follows. The second section discusses the materials and methods, including experimental details of the sample fabrication, structural and spectral characterization, and numerical modelling of the system for verification. The third section shows the experimental and computational results and their discussion. The last section concludes the results and provides insight for possible future applications.   
\section{Materials and methods}
\subsection{Experimental Details}
The sample contains a layer of anisotropic-nanoporous alumina on the aluminium substrate. To fabricate the sample, we used an aluminium sheet of 99.999\% purity obtained from the Alfa Aesar. The sheet was cleaned in acetone. To remove the surface roughness of the sheet, it was electropolished in the ice-cold mixture of ethanol and perchloric acid. The electropolishing gave a smooth and mirror shiny surface of the aluminium. The electropolished aluminium sheet was anodized in a 0.3 M oxalic acid solution at 40 V and 0$^{\circ}$ C for 10 minutes. Thin nanoporous alumina was etched off in a mixture of chromic acid and phosphoric acid solutions at ambient conditions and then cleaned with DI water and dried off.  Second anodization was carried out again in the oxalic acid solution by keeping all parameters the same as for the first anodization for 15 minutes. After the second anodization, the sample was washed in DI water and dried for further characterization and use. The obtained sample had an optically transparent anisotropic-nanoporous alumina layer on the aluminium substrate. We have already reported more details of the anodization setup and processes elsewhere~\cite{pradhan2021thin}. Topographical and structural characterization was carried out using Zeiss's field emission scanning electron microscope (FESEM) Sigma. The spectral measurements were conducted using a Fourier transform infrared (FTIR) spectrometer (Agilent, Cary 660). The angle and polarization-dependent measurements (for both s- and p-polarized) were carried out using a commercially available adapter (Harrick/Sea Gull) connected to the spectrometer. We have used a highly reflecting aluminium mirror to normalize the reflectance from the fabricated samples. The base aluminium layer at the bottom prohibits any IR light from passing through the sample resulting in zero light transmission.  
\subsection{Computational details}
To verify our experimental results, we performed a numerical simulation study in the finite element method-based commercial software COMSOL Muliphysics$\textsuperscript{\textregistered}$.  For the simulation, a rectangular domain was considered. Unwanted reflections were avoided using perfectly matched layers (PMLs) along the direction of propagations. Wave-port boundary conditions are applied to excite the electromagnetic wave from the top of the simulation unit.  Then S-parameters were used to calculate the wavelength-dependent reflectivity. Since the nanopores were void and there was no filling, the pore material was assumed to be air with unit permittivity and permeability. We homogenized the nanoporous alumina using the Bruggeman homogenization method~\cite{mackay2012bruggeman} and obtained the effective anisotropic permittivity. The system has uniaxial anisotropy of  $\varepsilon_x = \varepsilon_y \ne \varepsilon_z$ where nanopores are oriented along $z$-direction. The optical indices of alumina and aluminium used for the homogenization process were obtained from the experimental data reported in  Ref.~\cite{kischkat2012mid}, and \cite{rakic1995algorithm} respectively. The calculated effective anisotropic permittivity of nanoporous alumina was used in the numerical simulation to calculate the reflectivity of the structure.  All the geometrical parameters such as the thickness of nanoporous alumina and the fill fraction of the nanopores used in the numerical simulation are the actual parameters of the fabricated sample mentioned in the experimental section. The simulations were performed for both the p- and s-polarized light with varying incident angles.  
\section{Results and Discussions}
\begin{figure}
\centering
\includegraphics[width=1\textwidth]{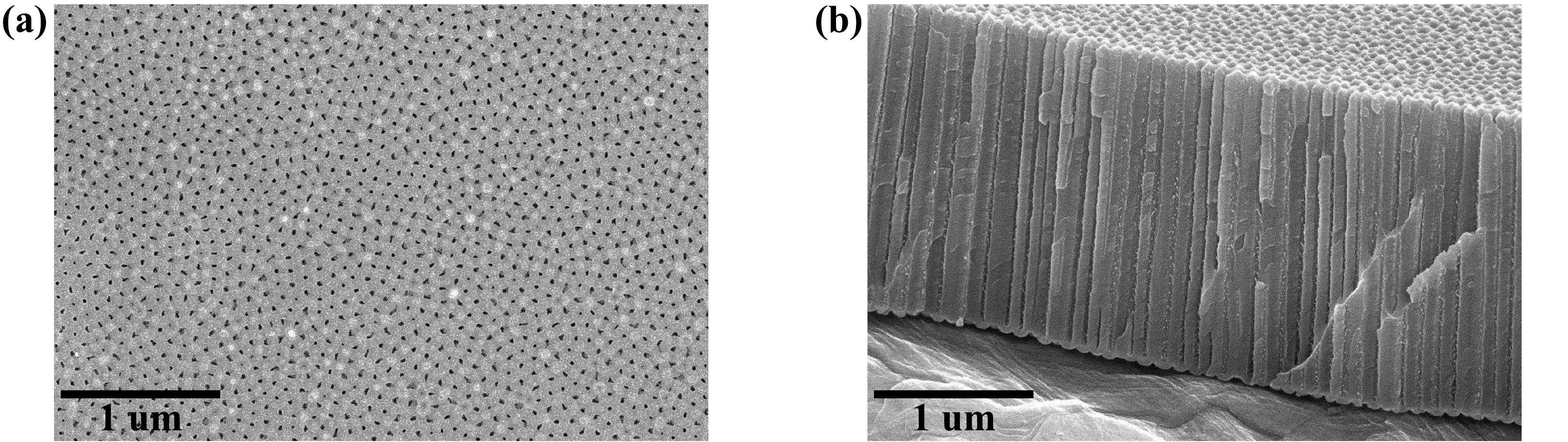}
\caption{The field emission scanning electron microscope (FESEM) image of the anisotropic nanoporous alumina on aluminium substrate. (a) Top surface shows the nanopores and (b) cross-section with aluminium base shows the straight and uniform varying nature of the nanopores.}
\label{fig:SEM_image}
\end{figure}

The FESEM image of the anisotropic-nanoporous alumina on the aluminium substrate is shown in Fig.~\ref{fig:SEM_image}. The surface topography of the nanoporous alumina is shown in Fig.~\ref{fig:SEM_image}(a). The average diameter of the nanopores is 20~nm, and the interpore distance is 90~nm, giving a 0.04 fill fraction of the nanopores. By default the nanopores are empty (filled with air). A clear cross-sectional view of the system is shown in Fig.~\ref{fig:SEM_image}(b). The reported anisotropic nanoporous slab on the aluminium substrate was 1.89~$\mu$m. It could be noted that the nanopores are very straight and uniform that giving a uniaxial anisotropic nature to the slab at the sub-wavelength scale. If distribution of the nanopores are in the $xy-$plane and orientated along the $z-$direction then this system has an optical anisotropy of type $\varepsilon_x = \varepsilon_y \ne \varepsilon_z$. This anisotropic behaviour of the nanoporous alumina slab can be geometrically controlled by varying the size of the nanopore and interpore distance by controlling the chemical and physical parameters while fabrication. 

\begin{figure}
\centering
\includegraphics[width=0.9\textwidth]{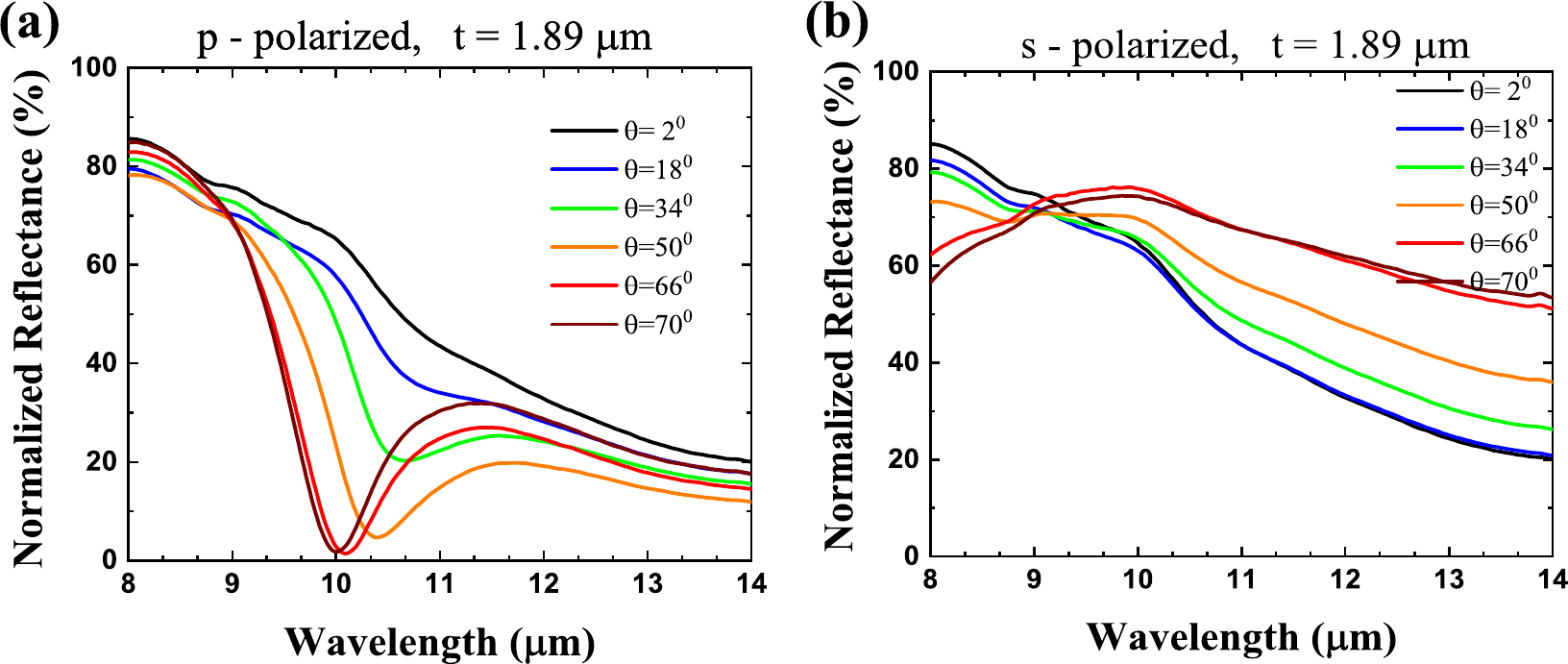}
\caption{Fourier transform infrared microscope (FTIR) measured reflectance of the anisotropic nanoporous alumina on aluminium substrate for (a) p-polarized light and (b) s-polarized light at different light incident angles, $\theta$, 0$^\circ$, 18$^\circ$, 34$^\circ$, 50$^\circ$, 66$^\circ$ and 70$^\circ$.}
\label{fig:exp_R_vs_lda_with_theta}
\end{figure}

\begin{figure}
\centering
\includegraphics[width=0.9\textwidth]{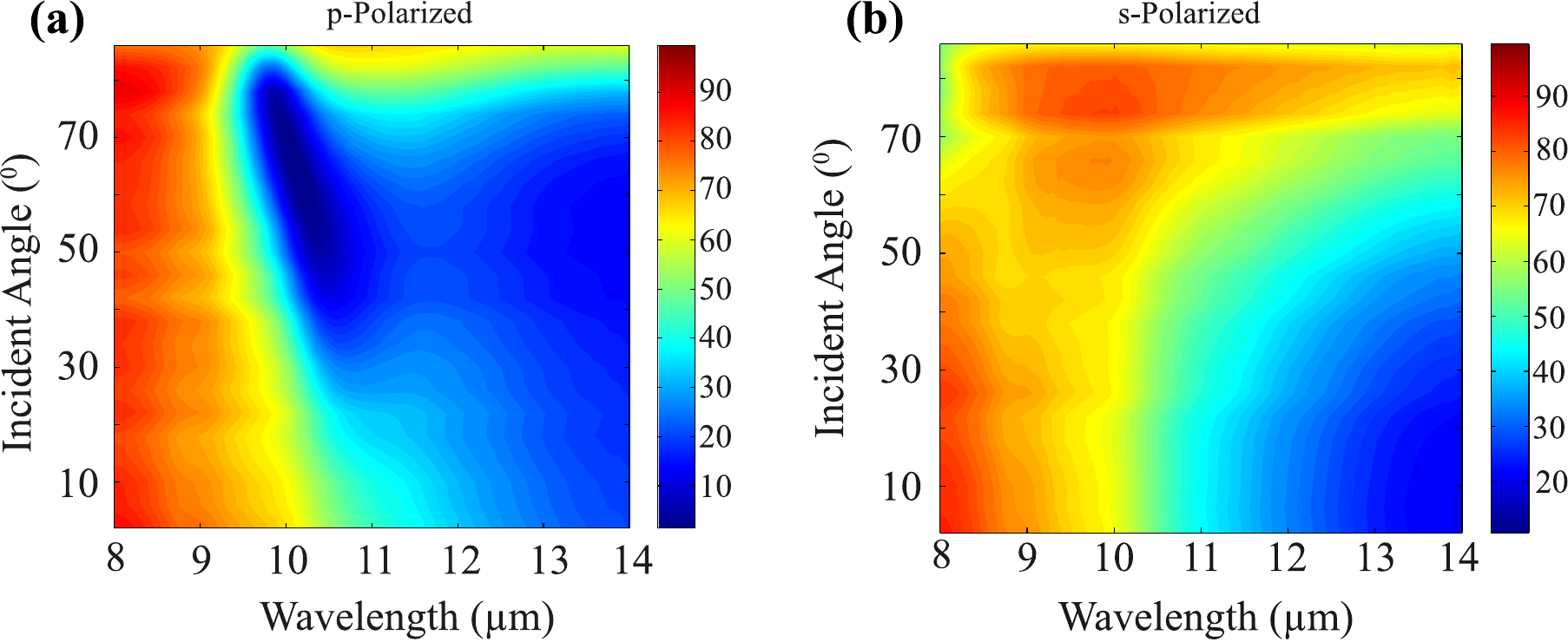}
\caption{Fourier transform infrared microscope (FTIR) measured reflection density plot for the anisotropic nanoporous alumina on aluminium system for (a) p-polarization and (b) s-polarization of the incident light.}
\label{fig:density_s_p_pol}
\end{figure}

Fig.~\ref{fig:exp_R_vs_lda_with_theta} shows the measured reflection for both the polarizations, p- and s-, of the incident light  at different fixed incident angles. As shown in Fig.~\ref{fig:exp_R_vs_lda_with_theta}(a), the  reflectance curve exhibits a near straight line behaviour for p-polarized light with angle of incident $\theta = 0^{\circ}$. However, with the increase of angle of incidence, a dip in the reflectance is observed. The dip hits a minimum at an angle, $\theta = 66^{\circ}$ at the wavelength $\lambda = 10.09~\mu$m. At this wavelength, the absorption goes as high as 98.62 $\%$ with minimal reflectance so that the sample behaves as a band-selective near  perfect absorber also. However, a different behaviour was observed from the sample when we measured the reflectivity for s-polarized light. The sample shows a gradual increase in the  reflectance with the incident angles(Fig.~\ref{fig:exp_R_vs_lda_with_theta}(b)). The results for s-polarized light are astonishingly different from that of p-polarized light. The observed dip in the reflectance for p-polarized light was expected to be a Berreman mode which was directly excited from the far-field. However, this Berreman mode was not present in the reflectance of s-polarised light.  These results contrast with the results for the light with s-polarized orientations where the reflectance increases with the angle of incidence. The extinction of the reflection for the p-polarized light shows the Berreman mode, which occurs at a lower wavelength where the dielectric permittivity tends to zero. This shows the potential use of the structures as a resonant emitter at mid-IR frequencies for an oblique angle of incidence. The measured reflection density plots with incident angle and wavelength are shown in Fig.~\ref{fig:density_s_p_pol}. We can see that for the p-polarized light; there is a region between the angle range of 70$^{\circ}$-60$^{\circ}$ and wavelength range 9.8~nm-10.5~nm, the reflection is minimum while for the same angle and wavelength range, the reflection is very high for the s-polarized light. This is the region where the Berreman modes exist for the p-polarized light. The minima positions for the 2$^{\circ}$ and 18$^{\circ}$ inclination of the incident light are not well defined from the shallow reflectance dips. The minimum reflectance and corresponding wavelengths are listed in Table~\ref{tab:table} for the light incident angles 34$^{\circ}$, 50$^{\circ}$, 66$^{\circ}$ and 70$^{\circ}$. As the incident angle increases, the reflectance minima shows a blueshift. Unlike the p-polarized light, for the s-polarized light, reflectance minima could not be defined. 

\begin{figure}
\centering
\includegraphics[width=0.45\textwidth]{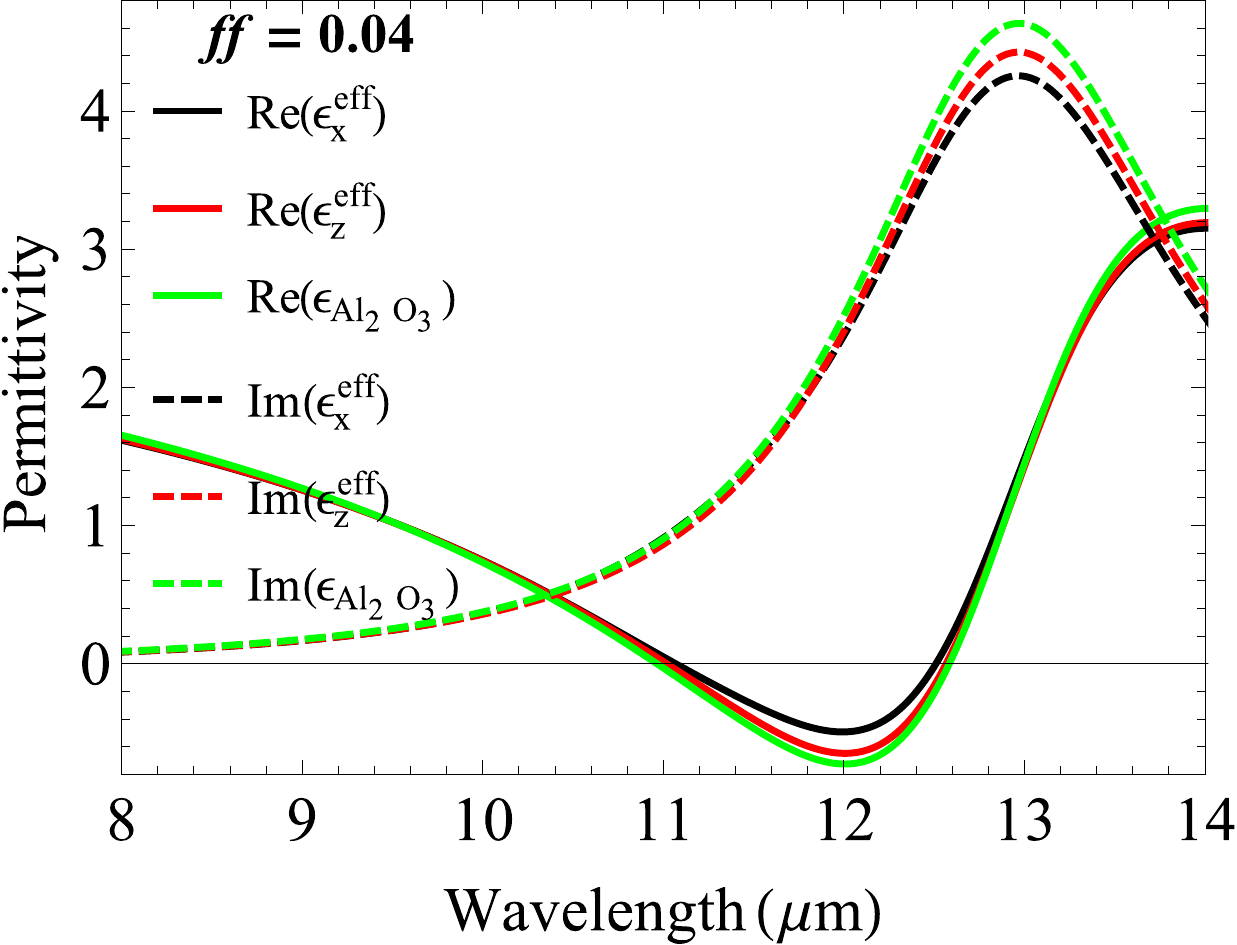}
\caption{The anisotropic effective permittivity of the nanopoaorous alumina obtained using the Bruggeman homogenization method at the pore fill fraction, $ff$ = 0.04. The experimental permittivity of host alumina ($\mathrm{Al_2O_3}$) and aluminium metal were used from Ref.[~\cite{kischkat2012mid}, and \cite{rakic1995algorithm}] respectively.}
\label{fig:effective_permittivity}
\end{figure}

The computed reflectance from the nanoporous alumina on aluminium structures using COMSOL Multiphysics shows similar characteristics as was measured experimentaly. The nanopores fill fraction is very small, $ff = 0.04$. The effective anisotropic permittivity of the nanoporous alumina calculated using the Bruggeman homogenization method is shown in Fig.~\ref{fig:effective_permittivity} at fill fractions 0.04. For comparison, we have plotted various components of the permittivity of the alumina host along with those of porous alumina. The nanoporous alumina displays small optical anisotropy because of the small fill fraction of the nanopores resulting in very similar dielectric properties compared to the pure alumina (Fig.~\ref{fig:effective_permittivity}). As the size of the nanopores could be very easily increased by widening the pores, and that could increase the fill fraction and hence the optical anisotropy. At higher fill fractions, the anisotropy would be higher and system might show different properties but that would be the subject of further study. However, here we concentrate only to verify the existence of the Berreman modes in our simple experimentally observed system at fill fraction of 0.04.   The real parts of the effective permittivity components $\mathrm{Re}(\varepsilon_x) = 0$ at wavelengths 11.06$~\mu$m and 12.51$~\mu$m, and $\mathrm{Re}(\varepsilon_z) = 0$  at wavelengths 11.01$~\mu$m and 12.57~$\mu$m respectively while the real part of the permittivity of the host alumina $\mathrm{Re}(\varepsilon_{\mathrm{Al_2O_3}}) = 0$ occurs at 10.96~$\mu$m and 12.58~$\mu$m. The maximum negative  values of real parts of effective $\varepsilon_x$, $\varepsilon_z$ and for host alumina $\varepsilon_{\mathrm{Al_2O_3}}$  occur at 11.99$~\mu$m, 12.00$~\mu$m and 12.00$~\mu$m respectively. We can observe the Reststrahlen band~\cite{zhao2010options} for the nanoporous alumina between wavelength 11.06~$\mu$m to 12.51~$\mu$m for both components $\varepsilon_x$ and $\varepsilon_z$ simultaneously where the real part is smaller than the imaginary part of the permittivity, resulting in a negative permittivity of the nanoporous alumina (Fig.~\ref{fig:effective_permittivity}). This behaviour is attributed to the phononic resonance bands of the nanoporous alumina with empty nanopores. 
\begin{table}
	\caption{Minima position in experimental and calculated reflectance for p-polarized light at various angle of incidence in nanoporous alumina on aluminium substrate. For the anisotropic nanoporous alumina thickness = 1.89~$\mu$m, $ff$ = 0.04, $\lambda |_{\mathrm{Re(\varepsilon_x) = 0}} = 11.06~\mu$m, $\lambda |_{\mathrm{Re(\varepsilon_z) = 0}} = 11.01~\mu$m, and for the isotropic host alumina $\lambda |_{\mathrm{Re(\varepsilon) = 0}} = 10.96~\mu$m.}
	\centering
	\begin{tabular}{ccccc} 
		\toprule
		\toprule	 
		\multirow{2}{*}{Incident angle ($^{\circ}$)} & \hspace{20mm} Measured  &  & \hspace{20mm} Computed  \\
				\cmidrule(r){2-3}
				\cmidrule(r){4-5}
		 	   	  & R$_{\mathrm{min}}$ (\%) & $\lambda_{\mathrm{R_{min}}}$ ($\mu$m) & R$_{\mathrm{min}}$ (\%) & $\lambda_{\mathrm{R_{min}}}$ ($\mu$m)\\
		\midrule
		34  	  & 20.26  & 10.65  & 20.12  & 10.64  \\ 
		50 	      & 4.65   & 10.39  & 4.64   & 10.37  \\ 
		66 	      & 1.38   & 10.09  &  0.75  & 9.92   \\ 
		70        & 1.75   & 10.01  &  1.10  & 9.80   \\ 
		\bottomrule
		\bottomrule
	\end{tabular}
	\label{tab:table}
\end{table}

\begin{figure}
\centering
\includegraphics[width=1\textwidth]{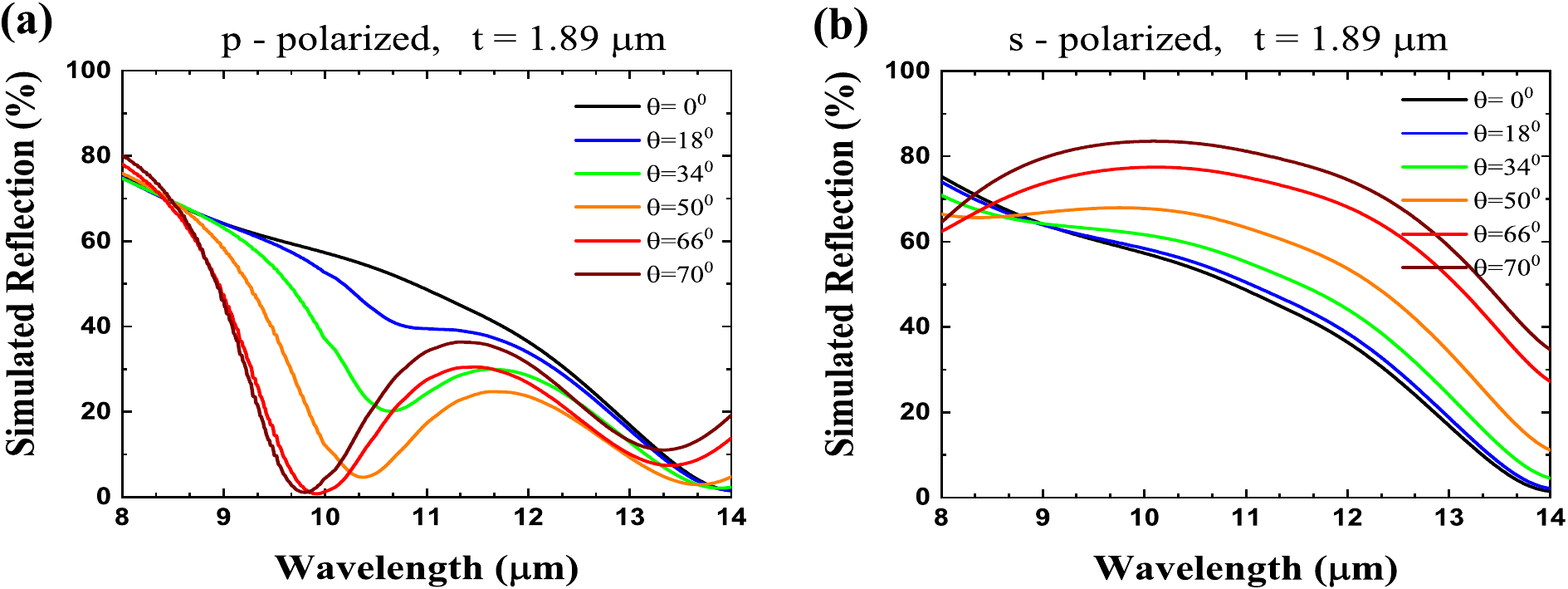}
\caption{Calculated reflectance of the nanoporous alumina/aluminium system for (a) p-polarization and (b) s-polarization at different light incident angles ($\theta$) 0$^\circ$, 18$^\circ$, 34$^\circ$, 50$^\circ$, 66$^\circ$ and 70$^\circ$.}
\label{fig:cal_R_vs_lda_with_theta}
\end{figure}

The computed reflectance from the nanoporous alumina/aluminium system for p-polarised and s-polarised light is shown in Fig.~\ref{fig:cal_R_vs_lda_with_theta}. We can see that there is similar behaviour of the reflectance spectra as were in the experimental case. We got almost exact results to compare to the experiment. The calculated reflectance minima and their corresponding wavelengths are also listed in  Table~\ref{tab:table} to compare with the experimental results. For instance at incident angle $66^{\circ}$, the leaky Berreman modes are excited at a resonant wavelength of 9.92~$\mu$m at which the refracted waves propagate within the porous alumina layer while the $\mathrm{Re}(\varepsilon_{\mathrm{Al_2O_3}}) = 0$ is at 10.96~$\mu$m. This reflectance minima is at the lower side of the wavelength where ENZ condition occurs~\cite{vassant2013electrical}. These Beerreman modes are excited directly by a p-polarized light incident from the free space as these modes are within the light cone.  At much smaller incident angles, the propagation distance for the refracted ray is insufficient that results in lower absorption inside the structures (Fig.~\ref{fig:cal_R_vs_lda_with_theta}(a)). At a higher incident angle, the refracted light propagates through the porous alumina leading to a high resonant absorption~\cite{chen2013trapping}. For s-polarization the reflection was higher in the same spectral range where p-polarization reflectance was lowest (Fig.~\ref{fig:cal_R_vs_lda_with_theta}(b)). It was showing the same behaviour as was observed experimentally. 

\begin{figure}
\centering
\includegraphics[width=1\textwidth]{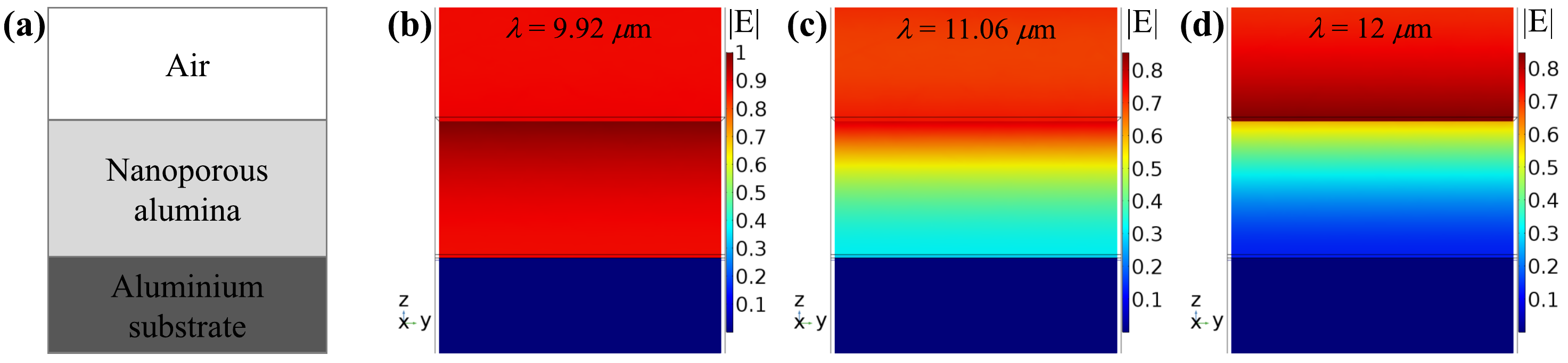}
\caption{(a) Sketch of the domains shows the air, alumina, and aluminium. Calculated normalized electric field plots for nanoporous alumina on aluminium substrate for p-polarized light at the incident angle 66$^{\circ}$ for (b) resonance wavelength 9.92~$\mu$m, (c) 11.06~$\mu$m where the real part of the effective permittivity becomes zero and (d) 12~$\mu$m where the real part of the effective permittivity is minimum.}
\label{fig:field_plots}
\end{figure}

The computed normalised electric field plots for the p-polarized light at the incident angle 66$^{\circ}$ for wavelengths 9.92$~\mu$m (at minimum reflection), 11.06$~\mu$m (at $\mathrm{Re}(\varepsilon_x) = 0$) and 12$~\mu$m (at maximum negative $\mathrm{Re}(\varepsilon_{\mathrm{Al_2O_3}})$) are shown in Fig.~\ref{fig:field_plots}. Fig.~\ref{fig:field_plots}(a) shows a sketch of trilayer asymmetric phononic systems that have domains of air, nanoporous alumina, and aluminium for which the electric field distributions are shown. As the reflectance minima occur at the 9.92~$\mu$m at the lower side (higher frequency) of the epsilon-zero position (11.06$~\mu$m), it is termed the Berreman mode. We can see that the field is highly concentrated inside the nanoporous alumina near the air-nanoporous alumina interface at this wavelength (Fig.~\ref{fig:field_plots}(b)) since the Berreman modes exist at the air-dielectric interface in an asymmetric system. The field at wavelength 11.06~$\mu$m where the real part effective permittivity becomes zero is not concentrated inside the nanoporous alumina but appears diffusive (Fig.~\ref{fig:field_plots}(c)). Since the maximum negative real part of the effective permittivity for the nanoporous alumina as well permittivity of host alumina is at 12~$\mu$m, therefore, at this wavelength there is no resonance and the electric field is reflected back into the air. At this wavelength, the nanoporous alumina itself almost behaves as a mirror (Fig.~\ref{fig:field_plots}(d)) because of metallic type character. The field inside the nanoporous alumina near the alumina-aluminium interface is very small. This reflective property of the dielectric polar material could be used as single layer mirror.

\section{Conclusions}
In conclusion, we have experimentally shown the minimum reflection for the mid-IR p-polarized light in the nanoporous alumina on aluminium substrate, while for the same wavelength position, there is a maximum reflection for the s-polarized light. We have numerically reproduced these experimentally observed results. At the resonance position, near the air-dielectric interface, inside the dielectric medium, the field is confined. Light is reflective and diffusive at the wavelengths where real parts of the permittivity become zero and the minimum negative value. These modes are analysed as the Berreman mode in the air/uniaxial-anisotropic-nanoporous-alumina/aluminium trilayer asymmetric phononic system. The nanoporous alumina is a very simple and easily achievable system with a high grade that could be fabricated over a large scale. These modes in this system can be used for thermal management, sensing, terahertz generation, and passive radiative cooling devices. 
 
\section*{Acknowledgement}
The authors would like to thank Venu Gopal Achanta of the Department of Condensed Matter Physics \& Materials Science, Tata Institute of Fundamental Research, Mumbai, India, for the discussion and suggestions.



\bibliography{references}   
\bibliographystyle{unsrt}

\end{document}